# Magnetostatic effect on spin dynamics properties in antiferromagnetic Van der Waals material CrSBr

Hongyue Xu[1], Nan Jiang[1], Haoran Chen[1], Yi Chen[2], Tong Wu[1], Yongwei Cui[1], Yunzhuo Wu[1], Zhiyuan Sheng[1], Zeyuan Sun[1], Jia Xu[3], Qixi Mi[2], Shiwei Wu[1,4,5,7], Weichao Yu[4,5,6]*, Yizheng Wu[1,7,8]*

[1]*Department of Physics and State Key Laboratory of Surface Physics, Fudan University, Shanghai 200433, China*

[2]*School of Physical Science and Technology, ShanghaiTech University, Shanghai 201210, China*

[3]*Department of Physics, School of Physics and Telecommunication Engineering, Shaanxi University of Technology, Hanzhong 723001, China*

[4]*Institute for Nanoelctronic devices and Quantum computing, Fudan University, Shanghai 200433, China*

[5]*Zhangjiang Fudan International Innovation Center, Fudan University, Shanghai 201210, China*

[6]*Shanghai Branch, Hefei National Laboratory, Shanghai 201315, China*

[7]*Shanghai Research Center for Quantum Sciences, Shanghai 201315, China*

[8]*Shanghai Key Laboratory of Metasurfaces for Light Manipulation, Fudan University, Shanghai 200433, China*

**Abstract**:

Van der Waals (vdW) antiferromagnets are exceptional platforms for exploring the spin dynamics of antiferromagnetic materials owing to their weak interlayer exchange coupling. In this study, we examined the antiferromagnetic resonance spectra of anisotropic Van der Waals antiferromagnet CrSBr. In addition to the ordinary resonance modes, we observed a dipolar spin wave mode when the microwave field was oriented perpendicular to the in-plane easy axis of CrSBr. Furthermore, our results uncovered a pronounced dependency of various resonant modes on the orientation of the microwave field, which is pivotal for the accurate determination of exchange coupling constants. Numerical simulations have elucidated this orientation dependence of spin dynamics arises from the magnetostatic effect. This discovery underscores the previously underappreciated significance of dipolar interactions in shaping the dynamical properties of two-dimensional AFM materials, thereby enhancing our understanding of the intrinsic dynamic properties of vdW magnets.

Van der Waals magnetic materials have drawn considerable research attention subsequent to the discovery of long-range magnetic ordering in compounds like $CrI_3$ [1], $Cr_2Ge_2Te_6$ [2], and $Fe_3GeTe_2$ [3]. These materials hold the promise of novel quantum phenomena and applications in next-generation devices [4-9]. The dynamic properties and spin wave phenomena of van der Waals antiferromagnets have garnered particular interest due to their zero net magnetization and high-speed operation capabilities [8-13]. Nevertheless, experimental investigation on the magnetization precession of antiferromagnets remains a substantial challenge due to their terahertz-frequency precession. Layered A-type 2D antiferromagnetic materials provide an outstanding platform due to their weak interlayer exchange coupling, leading to gigahertz resonance frequencies [14-17].

Previous studies on 2D antiferromagnets have revealed numerous phenomena that align well with theoretical predictions [15-26]. Techniques like broadband microwave absorption spectroscopy and inelastic neutron scattering have detected acoustic and optical resonance modes in these materials [15-18, 21, 24]. These modes can exhibit strong magnon-magnon coupling when rotational symmetry is broken [15-18,21]. Various spin dynamic phenomena, including nonlinear damping, standing spin waves, and magnon-photon coupling in antiferromagnetic-superconducting cavity systems, have been explored using materials like $CrCl_3$ [19-20, 22].

CrSBr is a typical air-stable A-type van der Waals antiferromagnetic material with a Néel temperature of 132 K [27-29]. CrSBr exhibits promising electrical, optical, and magnetic properties for spintronics research, with potential applications in spin logic and magnonic devices [7, 30-33]. The atomic structure and magnetic anisotropy of CrSBr are illustrated in Figure 1a. Atom layers are stacked along the *c* axis, exhibiting antiferromagnetic exchange coupling [28]. In terms of magnetic anisotropy, the *c* axis is the hard axis, and the *b* axis is the easy axis, while the *a* axis is the intermediate axis [28]. Due to its weak interlayer exchange coupling and in-plane anisotropic field, CrSBr displays rich spin dynamic phenomena [16, 23, 28-29]. Although the weak exchange coupling provides convenience for studying spin dynamics, it also makes exchange coupling comparable to dipolar interactions. Therefore, the influence of dipolar interactions on the spin dynamics of Van der Waals magnetic materials requires further exploration [14, 34].

We systematically investigated spin dynamics of CrSBr under varying orientations of both the microwave field and the magnetic field relative to the crystal axes, attaining insights into the interplay between dipolar interactions and antiferromagnetic resonance modes. Our discoveries revealed a dipolar mode with the microwave field perpendicular to the in-plane easy axis of CrSBr. Meanwhile, we found that the resonance frequency of the optical mode can be adjusted by the microwave field direction. Furthermore, with higher magnetic fields, the ferromagnetic resonance frequency varied with the microwave field direction due to the magnetostatic effect. To further understand our experimental results, we conducted numerical simulations demonstrating good consistency with the experimental data. Our work not only enhances the understanding of spin dynamics in CrSBr but also broadens the knowledge regarding dipolar interactions in 2D antiferromagnetic materials.

Figure 1c, d show the setups for measuring the antiferromagnetic resonance absorption spectra of CrSBr. The sample, which is a 4 mm × 0.8 mm needle-like crystal with its long edge along the $a$ axis, is positioned on a coplanar waveguide (CPW). In Figure 1c, the microwave magnetic field is parallel to the $a$ axis, while in Figure 1d, it is perpendicular to the $a$ axis. Thus, the effect of microwave orientation on the antiferromagnetic resonance can be experimentally explored. θ and φ represent the orientation of the external magnetic field relative to the $c$ axis and $a$ axis, respectively. The antiferromagnetic resonance is excited by the microwave current through the CPW. The microwave field predominantly lies in-plane on the signal line and out of plane in the CPW gap. The microwave transmission $\Delta S_{21}$ is measured while varying field strengths and orientations, and further normalized to the sample area on the signal line.

We initially measured the absorption spectra as a function of temperature within the zero magnetic field with $m_1$ and $m_2$ aligned along the $b$ axis (Figure 1e, f). The resonance modes are depicted in Figure 1b. CrSBr exhibits two resonance modes at zero magnetic field due to its hard axes along the $a$ and $c$ axes. The in-plane (IP) mode corresponds to the total magnetic moment $m$ oscillating along the $a$ axis, while the out-of-plane (OOP) mode corresponds to $m$ oscillating along the $c$ axis [38-40]. In the $h_{rf} \perp a$ configuration, the IP mode is prohibited by symmetry but can be excited in the $h_{rf} || a$ configuration [15-17,35,36]. The OOP mode persists in both configurations due to the perpendicular microwave field in the CPW gap. An extra dipolar mode emerges below 115 K for $h_{rf}||a$ (red arrows in Fig. 1f) exhibits a temperature dependence similar to that of the in-plane mode.

At 5 K, we performed field-dependent microwave absorption measurements with $H$ oriented along the three crystal axes (Figure 2). The red dashed lines in Figure 2 represent the field-dependent dispersion relations derived from the coupled Landau-Lifshitz (LL) equation [37]:

$$\frac{dm_1}{dt} = -\mu_0\gamma m_1 \times (H - H_E m_2 - H_a(m_1 \cdot a)a - H_c(m_1 \cdot c)c - \frac{M_s}{2}(m_{1z} + m_{2z})z),$$

$$\frac{dm_2}{dt} = -\mu_0\gamma m_2 \times (H - H_E m_1 - H_a(m_2 \cdot a)a - H_c(m_2 \cdot c)c - \frac{M_s}{2}(m_{1z} + m_{2z})z). \quad (1)$$

Here, $\mu_0$ is the permeability constant, $\gamma$ is the gyromagnetic ratio, $H$ is the external field, $H_E$ is the interlayer exchange field strength, $H_c$ is the out-of-plane anisotropy field, $H_a$ is the in-plane uniaxial anisotropy field, and $M_s$ is the saturation magnetization. The term $-\frac{M_s}{2}(m_{1z} + m_{2z})$ refers to the demagnetization field due to the total magnetization component perpendicular to the sample plane, which is different from previous literature [15-18]. With vanishing $M_s$, our formula matches previous calculations excluding the demagnetization field [16].

When $H$ is perpendicular to the easy-axis, both $m_1$ and $m_2$ tilt towards the field direction, forming optical and acoustic modes (Figure 2g, h). Along the $a$ axis, the optical mode corresponds to the total magnetic moment $m$ ($m = m_1 + m_2$) oscillating along the $a$ axis and the acoustic mode corresponds to the oscillation around the $a$ axis. The optical mode can only be excited in the $h_{rf}||a$ configuration due to the symmetry restriction, while the acoustic mode can be excited in both configurations [15-17,35,36]. The resonance frequencies of both optical mode and acoustic mode are given by $\omega_O = \mu_0\gamma\sqrt{\frac{H_c((2H_E+H_a)^2-H^2)}{2H_E+H_a}}$ and $\omega_A = \mu_0\gamma\frac{\sqrt{(H_c+2H_E+M_s)(H_a(2H_E+H_a)^2+H^2(2H_E-H_a))}}{2H_E+H_a}$, respectively. Subsequently, the frequency of optical mode decreases with increasing $H$, while the frequency of the acoustic mode continues to rise, and they cross at 0.58 T. Above 1 T, both $m_1$ and $m_2$ align parallel to the magnetic field, transitioning the acoustic mode into the ferromagnetic resonance (FMR) mode. The resonant frequency ($\omega_a$) increases almost linearly with $H$, as shown in Figure 2a, d, and is given by $\omega_a = \mu_0\gamma\sqrt{(H-H_a)(H-H_a+H_c+M_s)}$ increases almost linearly with $H$, as shown in Figure 2a, d.

When $H$ is oriented along the $b$ axis and is below the spin-flip field of 0.33 T, $m_1$ and $m_2$ maintain antiparallel alignment. This arrangement gives rise to right-handed IP and left-handed OOP modes. For the IP mode, the precession amplitude of $m_2$ along $+b$ axis increases as $H$ increase along $+b$ axis, resulting in an elliptical precession of $m$ with the major axis along the $a$ axis (Figure 2i). The resonance frequency increases with $H$, and is given by $\omega_{RH} = \mu_0\gamma\sqrt{H^2 + H_cH_E + H_a\left(H_c + H_E + \frac{M_s}{2}\right) + \sqrt{H^2(H_c + H_a)(H_c + H_a + 4H_E + M_s) + (H_cH_E - H_aH_E - \frac{M_s}{2}H_a)^2}}$, as shown in Figure 2b, e. For the OOP mode, the precession amplitude of $m_1$ along $-b$ axis increases as $H$ increase along $+b$ axis. Thus, $m$ forms an elliptical shape with the major axis along the $c$ axis (Figure 2j). The resonance frequency decreases with $H$ and is given by $\omega_{LH} = \mu_0\gamma\sqrt{H^2 + H_cH_E + H_a\left(H_c + H_E + \frac{M_s}{2}\right) - \sqrt{H^2(H_c + H_a)(H_c + H_a + 4H_E + M_s) + (H_cH_E - H_aH_E - \frac{M_s}{2}H_a)^2}}$ (Figure 2b, e). When $H$ surpasses 0.33 T, a spin-flop transition occurs. Beyond the saturation field (0.35 T for our sample), the derived resonance frequency of FMR mode is given by $\omega_b = \mu_0\gamma\sqrt{(H + H_a)(H + H_c + M_s)}$.

For $H$ along the $c$ axis, the optical mode corresponds to $m$ oscillating along the $c$ axis, while the acoustic mode corresponds to $m$ rotating around the $c$ axis. The acoustic mode exhibits a higher frequency $\omega_A = \mu_0\gamma\frac{\sqrt{(H_a + 2H_E)(H_c(H_c + 2H_E + M_s)^2 - H^2(H_c - 2H_E))}}{H_c + 2H_E + M_s}$, while the optical mode displays a lower frequency $\omega_O = \mu_0\gamma\sqrt{\frac{H_a((2H_E + H_c + M_s)^2 - H^2)}{2H_E + H_c + M_s}}$. Both frequencies decrease with increasing field until saturation (Figure 2c, f). When $H$ is larger than $H_c + 2H_E + M_s$, the FMR mode can be observed. The FMR mode with $H||c$ has the resonance frequency $\omega_c = \mu_0\gamma\sqrt{(H - H_c - M_s)(H - H_c - M_s + H_a)}$. Through fitting, we obtained the magnetic parameters $\frac{\gamma}{2\pi} = 28.9\ GHz/T$, $\mu_0H_E = 0.35\ T$, $\mu_0H_a = 0.38\ T$, $\mu_0H_c = 1.2\ T$, $\mu_0M_s = 0.12\ T$. The calculated curves, described by the red dashed lines, show good agreements with experimental data. Comparing with the fitted parameters in Ref. 16, the exchange field $H_E$ is 13% smaller, $H_a$ and $H_c + M_s$ are similar.

When comparing the microwave absorption spectra measured with $h_{rf}||a$ and $h_{rf} \perp a$, we observed the following: i) When $H||a$, the dipolar mode exhibits a similar dispersion relation to the optical mode, and those modes are only excited in the $h_{rf}||a$ configuration (Figure 2a) [13-16]. ii)

For $H||b$, by comparing with the fitting curves, we can clearly see that the resonance frequency of the in-plane mode in the $h_{rf}||a$ configuration is higher than that in the $h_{rf} \perp a$ configuration, as indicated by purple arrows in Figure 2b and 2e. Such differences in resonance modes, attributed to the varying orientations of $h_{rf}$, can lead to inaccurate fitting if the analysis is based solely on data from a single configuration. Additionally, in the $h_{rf}||a$ configuration, the dipolar mode shows a similar dispersion relationship to the in-plane mode. iii) When $H||c$, the acoustic mode can be excited in the $h_{rf} \perp a$ configuration, which align with Eq. 1. In the $h_{rf}||a$ configuration, the frequency of the acoustic mode is significantly higher than that in the $h_{rf} \perp a$ configuration. The dipolar mode can also be observed, as depicted in Figure 2c. In all three $H$ directions, the dipolar mode is observable in the $h_{rf}||a$ configuration. In fact, the dipolar mode, while exists, has been overlooked in the data presented in Ref. 16. The dispersion relation of the dipolar mode varies with the $H$ direction, but it consistently resembles the uniform resonance mode, where the component of the total magnetic moment oscillates along the $a$ axis. For $H$ larger than the saturation field, when $H||h_{rf}$, the resonance frequency is lower, and the linewidth is narrower than when $H \perp h_{rf}$.

To further investigate the impact of the $h_{rf}$ direction, we conducted additional microwave absorption experiments, varying the $h_{rf}$ direction by positioning the sample on the CPW at different angles $\phi$ between the $a$ axis and the CPW signal line. Figure 3a depicts the schematic diagram of measurements under varying orientations of $h_{rf}$ Figure 3b presents the resonance spectra measured at zero field. The lower-frequency out-of-plane mode remains largely independent of $\phi$, whereas the higher-frequency in-plane mode near 34 GHz increases continuously with $\phi$. An additional peak, labeled 2* at 37 GHz at $\phi = 90°$ corresponds to the dipolar mode indicated by the red arrow in Figure 2a. Figure 3c-e shows the measured spectra with $H||a$ at various field strengths above 0.8 T, which is above the crossing field of 0.58 T between the optical and acoustic modes. Thus, the low-frequency mode is the optical mode, and the high-frequency mode is the acoustic mode. In Figure 3c, as $\phi$ increases, the frequency of the optical mode increases, while that of the acoustic mode decreases. For $H$ above 1 T, the acoustic mode transitions into the FMR mode. Thus, the results in Figure 3d, e with the $\mu_0 H$ =1.2 T and 1.5 T show that the frequency of FMR mode decreases while $\phi$ increasing.

Figure 3f-h display the representative absorption spectra measured with $H||b$. Figure 3f shows the results measured at a field of 0.2 T. Two distinct resonance modes are observable: a higher-frequency in-plane mode and a lower-frequency out-of-plane mode. The frequencies of both modes increase with $\phi$. Note that a dipolar mode is observable at ~39 GHz in the spectra with $\phi = 90°$, as indicated by the arrow labeled as 2*. In Figure 3g-h, where $H$ exceeds the spin-flop field, only the FMR mode can be observed, and its resonance frequency increases with $\phi$.

Figure 3i-k shows the resonance spectra with $H||c$. Since $H$ is not sufficiently strong to saturate the magnetic moments along the $c$ axis, two distinct resonance modes are observed, with the acoustic mode exhibiting a higher frequency than the optical mode. The frequency of the acoustic mode increases with $\phi$. Additionally, the dipolar mode is observed at $\phi = 90°$. The optical mode frequency also increases with $\phi$. In general, regardless of the specific resonance mode, when the total magnetic moment $m$ oscillates along the $a$ axis, the resonance frequency increases with $\phi$. Conversely, when $m$ oscillates along the $c$ axis, the resonance frequency decreases with the increase in $\phi$.

To further elucidate the origin of the strong influence of $h_{rf}$ orientation on the resonance frequency in CrSBr, we conducted numerical simulations using the Micromagnetics Module and Radio Frequency module within the COMSOL® software framework [41, 42]. Supplementary Information VI provides detailed descriptions of the simulation parameters and setup. Figure 4a, b illustrate the simulation configurations replicating the experimental setups in Figure 1a, b, representing the configurations with $h_{rf}||a$ and $h_{rf} \perp a$, respectively. The sample is depicted as a purple cuboid measuring 4 mm × 0.8 mm × 0.1 mm, and the yellow area represents the signal line measuring 7 mm × 0.36 mm × 0.3 mm, which is used for transmitting microwave current. The long edge of the sample corresponds to the $a$ axis. In the simulations, we first allow the magnetic moments to reach equilibrium under $H$, after which $h_{rf}$ excites AFM resonance via the antenna. Figure 4c-f illustrate the variation of the normalized mean precession amplitude with frequency for different orientations of $H$ and $h_{rf}$. The insets in these figures show the distribution of the resonance intensity at the corresponding resonance peak frequencies.

Figure 4c compares the simulated in-plane modes under the two configurations at zero field.

When $h_{rf} \perp a$, the in-plane mode is weakly excited by the out-of-plane component of microwave field (peak 1 in Figure 4c). When $h_{rf}||a$, the in-plane mode is directly excited by the in-plane component of microwave field. The non-uniformity of the microwave field induces a non-uniform precession of magnetic moments, leading to the formation of magnetostatic waves. This dipolar mode exhibits a dispersion relationship similar to that of the resonance modes with oscillation along $h_{rf}$, with a slightly higher frequency, as marked by peak 2* in Figure 4c. Figure 4d-f compare the resonance frequencies obtained under the two configurations when a substantial magnetic field is applied along the *a*, *b*, and *c* axes, a scenario that is analogous to what is observed in ferromagnets. According to our simulation, when $H||a$ (Fig. 4d), the configuration with $h_{rf} \perp a$ excites a uniform FMR mode and magnetostatic surface wave (MSSW) modes [43, 44]. Conversely, when $h_{rf}||a$, the out-of-plane component of RF field can also excite a weak FMR mode. Consequently, the resonance frequency in the $h_{rf}||a$ configuration is lower than that in the $h_{rf} \perp a$ configuration. When $H||b$, the backward volume magnetostatic spin wave (BVMSW) modes are observed in the $h_{rf} \perp a$ configuration (Figure 4e) [43, 44]. In contrast, in the $h_{rf}||a$ configuration, a uniform FMR mode is present. Consequently, the resonance frequency in the $h_{rf}||a$ configuration is larger than that in the $h_{rf} \perp a$ configuration. When $H||c$, forward volume magnetostatic spin wave (FVMSW) modes are excited in both configurations [43, 44]. These simulation results align well with our experiential results, providing a clear picture of non-uniform antiferromagnetic resonance phenomena.

Our research results reveal that dipole interaction significantly influence the resonance frequency of CrSBr, leading to the emergence of magnetostatic waves. While previous studies have extensively explored the effect of magnetostatic waves in ferromagnetic systems [43-46], our work focuses on elucidating their impact on antiferromagnetic resonance. By changing the sample size and signal line width, we explored the relationship between the resonance frequency and the wavelength of the dipolar modes. As shown in supplementary Fig. S6, reducing the sample length and signal line width results in a decreased wavelength, which in turn increases the frequency of the dipolar modes. As evidenced by supplementary Fig. S7, we observed that, as the exchange field $H_E$ increases and the saturation magnetization $M_s$ decreases, the frequency differences between the dipolar mode and the in-plane mode decrease in the $h_{rf}||a$ configuration. Similarly, the frequency differences among the in-plane modes with the microwave field along different directions diminish with increasing $H_E$ and decreasing $M_s$. These observations underscore the significant role of dipolar interactions in the A-type two-dimensional antiferromagnetic materials with small exchange coupling fields and large saturation magnetization. Therefore, understanding and accounting for the impact of dipole interactions is crucial when studying resonances in such materials.

In our study, we conducted a thorough investigation into resonance modes within the van der Waals antiferromagnetic CrSBr across varying microwave field and magnetic field orientations. According to our results, resonant modes of CrSBr are significantly affected by the orientation of the microwave field, which is a point of divergence from previous literature[16], where the impact of the dipolar interaction on the dispersion relation was not as extensively considered. This effect plays a crucial role in accurately determining the exchange coupling. Notably, we discovered a dipolar mode with the microwave field perpendicular to the in-plane easy axis of CrSBr, which is attributed to the strong dipolar interaction of the sample. The frequency dependence of $h_{rf}$ and the existence of dipolar mode can be well described by the numerical simulation. The resonance frequency of dipolar mode increases as the wavelength decreases. Those effects are enhanced for materials with strong magnetization and weak exchange coupling, which emphasized the importance of dipolar interaction while studying the spin dynamics of the van der Waals antiferromagnetic materials. Our systematic antiferromagnetic resonance measurements and numerical simulation of CrSBr shows how the resonance frequency of antiferromagnets can be tuned by the inhomogeneous microwave field, which pave ways for a new method for manipulating and controlling spin dynamics in antiferromagnetic materials.

Acknowledgements:

The work was supported by the National Key Research and Development Program of China (Grant No. 2022YFA1403300), the National Natural Science Foundation of China (Grant No. 11974079, No. 12274083, No. 12434003, No. 12204107 and No. 12221004), the Shanghai Municipal Science and Technology Major Project (Grant No. 2019SHZDZX01), and the Shanghai Municipal Science and Technology Basic Research Project (Grant No. 22JC1400200 and No. 23dz2260100), the Innovation Program for Quantum Science and Technology (Grant No.

2024ZD0300103), Shanghai Science and Technology Committee (Grant No. 21JC1406200). W. Yu acknowledges fruitful discussions with Prof. Jiang Xiao.

Data Availability Statement

The data that support the findings of this study are available from the corresponding author upon reasonable request.

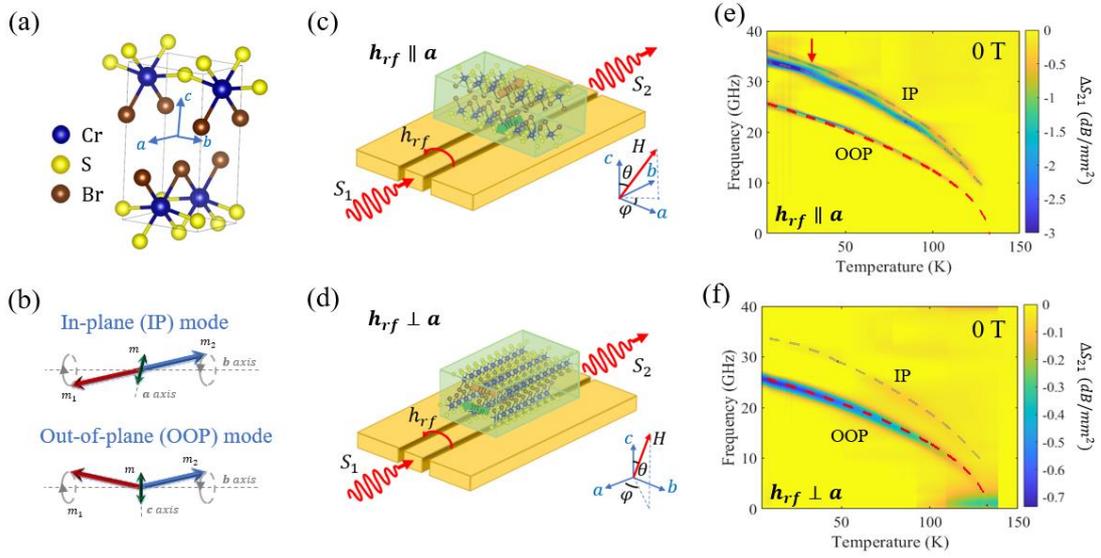

**Figure 1. Experiment setup and temperature dependence of resonance frequency.** (a) Atomic structure of CrSBr. (b) Two zero-field resonance eigenmodes of CrSBr. The red and bule arrows shows the procession of $m_1$ and $m_2$, and the green arrow shows the procession of total magnetic moment. (c-d) Experimental schematic of a CrSBr crystal mounted on a coplanar waveguide for microwave absorption measurements with (c) $h_{rf} \| a$ and (d) $h_{rf} \perp a$. (e-f) Temperature dependence of absorption spectra measured at 0 T with (e) $h_{rf} \| a$ and (f) $h_{rf} \perp a$. The microwave absorption intensity $\Delta S_{21}$ is the transmission rate normalized by the overlap area between the sample and the signal line.

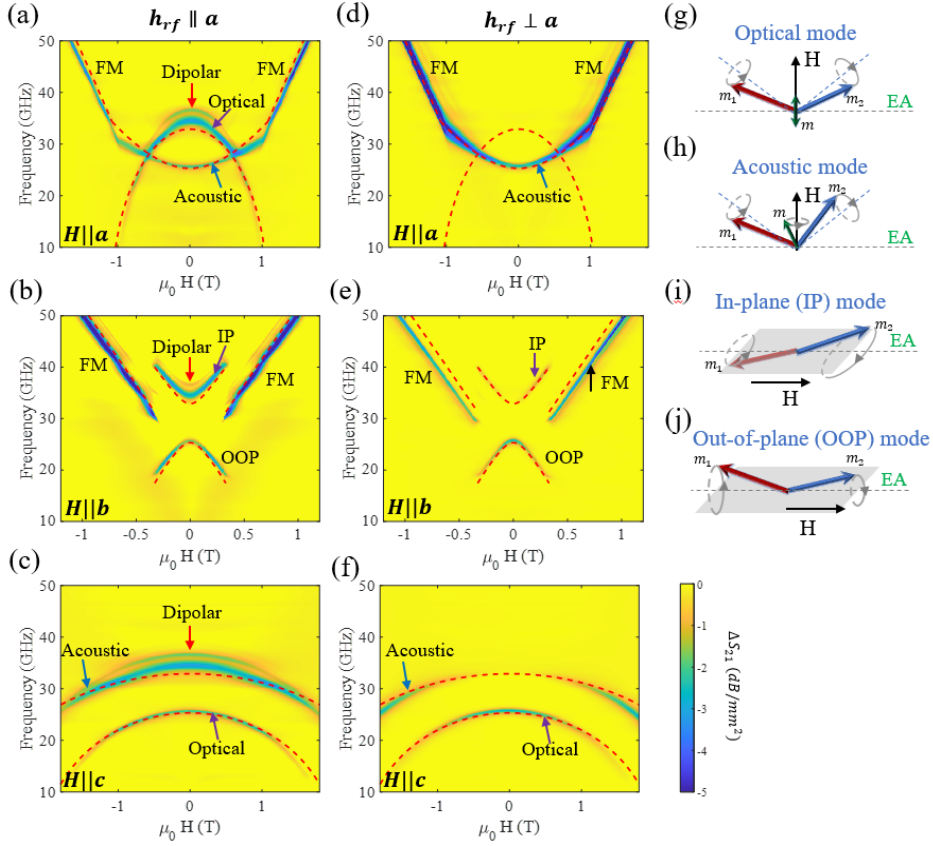

**Figure 2. Microwave absorption spectra measured at 5 K as a function of magnetic field applied along different axes with different microwave field direction.** (a-c) Absorption spectra measured with $h_{rf}||a$ for the field along three axes: (a) $H||a$, (b) $H||b$, (c) $H||c$. (d-f) Absorption spectra measured with $h_{rf} \perp a$ for the field along three axes: (d) $H||a$, (e) $H||b$, (f) $H||c$. The red dashed lines are the calculated curves based on the formula described in the text. (g) The diagram of optical mode and acoustic mode for the field perpendicular to the easy axis (***b***). (h) The diagram of in-plane mode and out-of-plane mode for the field parallel to the ***b*** axis.

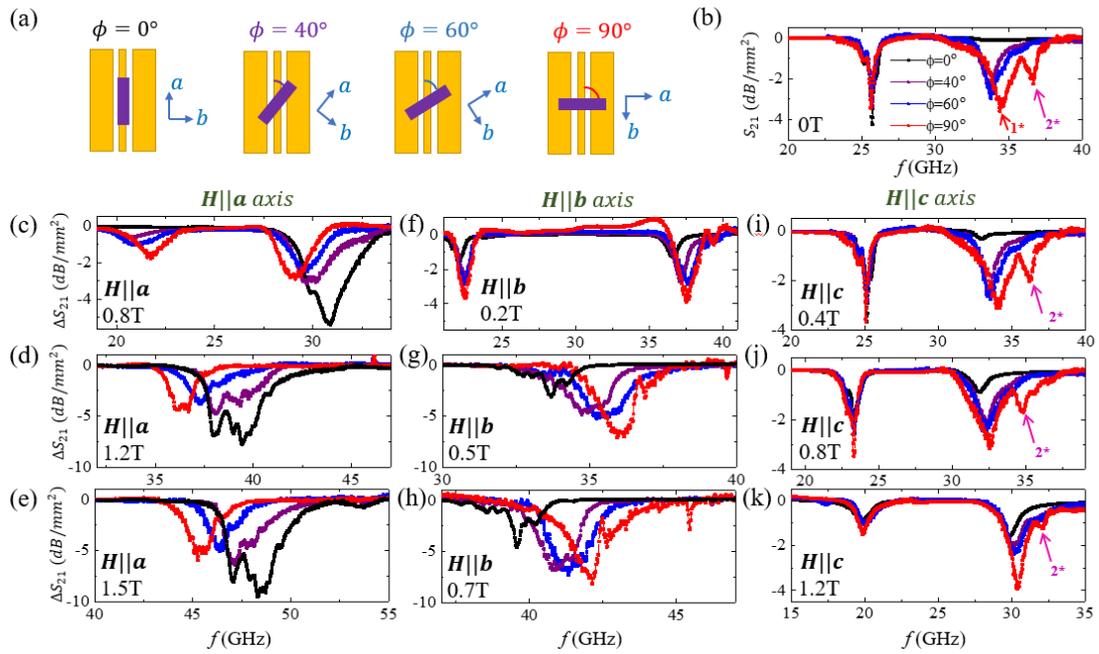

**Figure 3. Microwave field orientation-dependent absorption spectra in CrSBr.** (a) Schematic diagram illustrating the measurement geometry with different sample orientation $\phi$. (b) Absorption spectra measured at zero field. (c-k) Absorption spectra with different field strengths for the field along different crystalline axes: (c-e) $H||a$, (f-h) $H||b$, and (i-k) $H||c$.

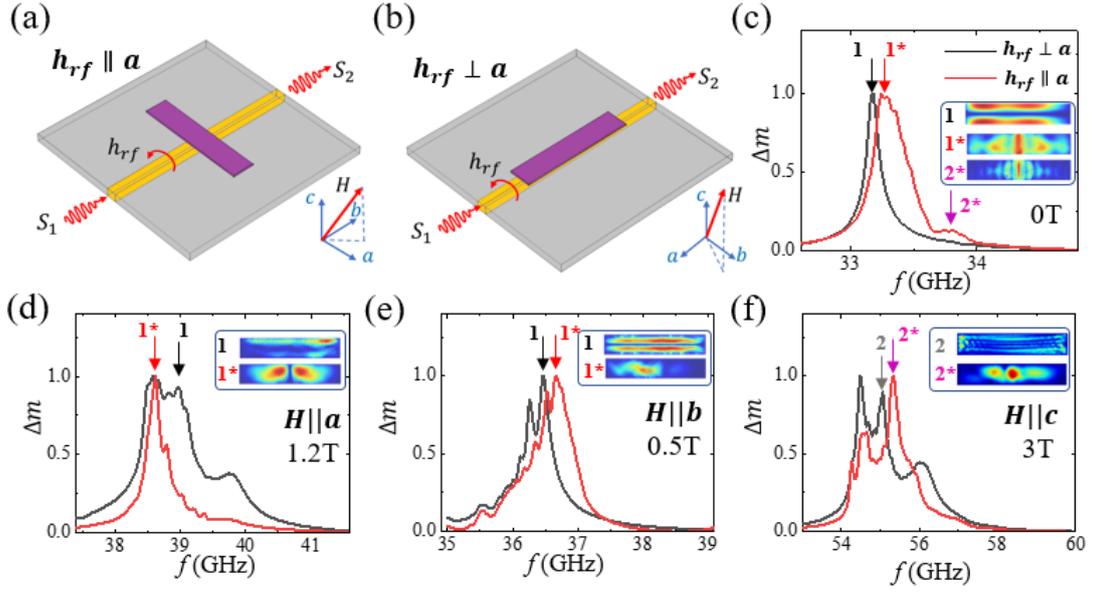

**Figure 4. Simulating the AFM resonance with different microwave field orientations.** (a-b) Simulation setup for the AFM resonance of CrSBr with (a) $h_{rf}||a$ and (b) $h_{rf} \perp a$. (c-f) The simulated dynamics spectra of CrSBr with two different $h_{rf}$ directions and different field directions: (c) $\mu_0 H$=0 T, (d) $\mu_0 H$=1.2 T, $H||a$, (e) $\mu_0 H$=0.5 T, $H||b$, (f) $\mu_0 H$=3 T, $H||c$. All the simulated spectra are normalized with the maxim peak. The insets in (c-f) are the spatial magnitude of the simulated resonance modes marked with the numbers in each figure.